\documentstyle[12pt]{article}
\begin{document}
\title{A Study of the $\pi B \rightarrow Y K $ reactions for   
Kaon Production in Heavy Ion Collisions\thanks{
{\it Talk given at Joint Japan-Australia Workshop on 
Quarks, Hadrons and Nuclei, Adelaide, South Australia, November 15-24, 1995 
(To appear in the conference proceedings)}
\newline ADP-96-4/T209
}}
\author{K. Tsushima$^1$\thanks{Supported by Australian Research Council}
S. W. Huang$^2$\thanks{Supported by GSI under contract No. T\"U F\"as T}
and Amand Faessler$^2$\\
$^1$Department of Physics and Mathematical Physics,\\
The University of Adelaide, Australia 5005\\
$^2$Institut f\"ur Theoretische Physik, Universit\"at T\"ubingen\\
Auf der Morgenstelle 14, 72076 T\"ubingen, F. R. Germany}
\date{\today}
\maketitle
\begin{abstract}
Parametrizations of total cross sections 
sufficient for all channels of the
$\pi B \rightarrow Y K$ 
reactions are completed using a resonance model.
As well as discussing the
$\pi N \rightarrow \Lambda K$
reactions, which were not presented in our previous 
publications, we present the differential cross section for 
$\pi N \rightarrow \Lambda K$. 
This report also aims at presenting supplementary discussions to our
previous work.
\end{abstract}
One of the main goals of studying heavy ion collisions is to 
determine the equation of state (EOS) of nuclear matter.
Because positive kaons ($K^+$) have a long mean free path inside the nucleus
they are suggested as a good probe for the reactions
occurring in the central region of the collisions \cite{sch}.
Indeed, theoretical studies show that the kaons produced in heavy ion 
collisions are sensitive to the EOS \cite{aic,lan}.  

Although one can point out many important ingredients
for the theoretical investigations of  kaon production 
in heavy ion collisions, the discussions presented here
are concerned with the elementary kaon production cross sections
necessary for microscopic calculations.

One of the purposes of this report is to complete the 
parametrizations of total cross sections
sufficient for all channels of 
the $\pi B \rightarrow Y K$ reactions by a resonance model 
\cite{tsu1,tsu2} ($B=N,\Delta$ and $Y=\Lambda,\Sigma$).
The results for the
$\pi N \rightarrow \Lambda K$ 
reactions which were not given in our previous publications \cite{tsu1,tsu2}
are given. 
Furthermore, supplementary disucussions to our previous 
work are presented. 

The microscopic transport models \cite{cas1} used for the calculations of 
kaon production in heavy ion collisions contain  
the following processes as the main collision terms:
$B_1 B_2 \rightarrow B_1 B_2$,\quad
$N N \rightarrow N N \pi$,\quad 
$N N \leftrightarrow N \Delta$,\quad
$\pi N \leftrightarrow \Delta$.\quad
Kaons are produced through the
$B_1 B_2 \rightarrow B_3 Y K$ and 
$\pi B \rightarrow Y K$ reactions.

For a given impact parameter $b$, the Lorentz-invariant 
differential kaon multiplicity in the microscopic calculations is given by
\begin{eqnarray}
E \frac{d^3N(b)}{d^3 p} &=& \sum_{B_1B_2} \int \left( E'
\frac{d^3\sigma _{B_1B_2 \rightarrow B_3YK} (\sqrt{s_{B_1B_2}})}{d^3 p'}  
/\sigma_{B_1B_2}^{tot}(\sqrt{s_{B_1B_2}}) \right)
[1-f({\bf r,p},t)] \frac{d\Omega_{3Y}}{4\pi}\qquad
\nonumber  \\ &  & 
+ \sum_{\pi B}  E''
\frac{d^3\sigma_{\pi B \rightarrow Y K} (\sqrt{s_{\pi B}})}{d^3 p''}/
\sigma_{\pi B}^{tot}(\sqrt{s_{\pi B}}).
\label{basic1}
\end{eqnarray}
\noindent
Here the primed and the double-primed quantities are in the
center-of-momentum (c.m.) frames of the 
two colliding baryons ($B_1B_2$) and pion-baryon ($\pi B$), respectively, 
while the unprimed quantities are those in the c.m. of the two nuclei. 
$\sigma_{B_1B_2}^{tot}(\sqrt{s_{B_1B_2}})$ and  $\sigma_{\pi B}^{tot}
(\sqrt{s_{\pi B} })$ are the total
cross sections 
as functions of the respective c.m. energies $\sqrt{s_{B_1B_2}}$ 
and $\sqrt{s_{\pi B}}$. The factor
$[1-f({\bf r,p},t)]$ stands for the Pauli blocking effects for the final
baryon $B_3$, and 
$\Omega_{3 Y}$ is the solid angle of the relative momentum
between the final baryon $B_3$ and hyperon $Y$.
The Lorentz-invariant double differential kaon production cross section
is obtained by integrating 
the kaon multiplicity Eq. (\ref{basic1}) 
over the impact parameter $b$ multiplied by the factor $2 \pi b$.
Eq.(\ref{basic1}) shows that the elementary 
kaon production cross sections are
directly related to the differential kaon yields. 
Thus, it is important to use correct  elementary 
kaon production cross sections for the microscopic investigations 
of kaon production in heavy ion collisions.

The elementary kaon production cross sections used for the 
microscopic calculations are 
usually taken from Randrup and Ko \cite{ran} in baryon-baryon collisions.
In pion-{\it nucleon} collisions 
they are usually taken from Cugnon and Lombard \cite{cug}. 
Our interest at the moment is with 
the latter case.
In the work of Cugnon and Lombard \cite{cug}  
total cross sections are parametrized by introducing an  
isospin-averaging procedure, and by using the 
limited experimental data 
assuming proton and neutron number $(N=Z)$ symmetry.
This means that the differential cross sections 
must be investigated using a more general procedure.
Furthermore, the reactions $\pi \Delta \rightarrow Y K$
cannot be investigated because no experimental data are available.
One of the motivations for this work is to study 
both the total and differential cross sections 
$\pi B \rightarrow Y K$ 
based on a microscopic model.
Below, we give a theoretical description of the 
$\pi N \rightarrow \Lambda K$ reactions which will be our main concern later.
The experimental data \cite{par} show that three
resonances $N(1650) (\frac{1}{2}^-)$, $N(1710) (\frac{1}{2}^+)$ and 
$N(1720) (\frac{3}{2}^+)$ make contributions to the   
$\pi N \rightarrow \Lambda K$
reactions.
In addition, $K^*(892)$ exchange is included as an effective
process which also simulates the other heavier $K^*$ meson effects.
The relevant processes are depicted in Fig. 1.
The interaction Lagrangians used are:
\begin{eqnarray}
{\cal L}_{\pi N N(1650)} &=&
-g_{\pi N N(1650)}
\left( \bar{N}(1650)  \vec\tau N \cdot \vec\phi
+ \bar{N} \vec\tau  N(1650) \cdot \vec\phi\,\, \right), 
\label{pinn1650}\\
{\cal L}_{\pi N N(1710)} &=&
-ig_{\pi N N(1710)}
\left( \bar{N}(1710) \gamma_5 \vec\tau N \cdot \vec\phi
+ \bar{N} \vec\tau \gamma_5 N(1710) \cdot \vec\phi\,\, \right), 
\label{pinn1710}\\
{\cal L}_{\pi N N(1720)} &=&
\frac{g_{\pi N N(1720)}}{m_\pi}
\left( \bar{N}^\mu(1720) \vec\tau N \cdot \partial_\mu \vec\phi
+ \bar{N} \vec\tau N^\mu(1720) \cdot \partial_\mu \vec\phi \,
\right),
\label{pinn1720}\\
{\cal L}_{K \Lambda N(1650)} &=&
-g_{K \Lambda  N(1650)}
\left(\bar{N}(1650) \Lambda K 
+\bar{K}\bar{\Lambda}  N(1650)  \,\, \right), 
\label{klan1650}\\
{\cal L}_{K \Lambda N(1710)} &=&
-ig_{K \Lambda  N(1710)}
\left( \bar{N}(1710) \gamma_5  \Lambda K 
+\bar{K}\bar{\Lambda} \gamma_5  N(1710)  \,\, \right), 
\label{klan1710}\\
{\cal L}_{K \Lambda N(1720)} &=&
\frac{g_{K \Lambda N(1720)}}{m_K}
\left( \bar{N}^{\mu}(1720)   \Lambda \partial _{\mu}  K 
+  (\partial_{\mu} \bar{K}) \bar{\Lambda} N^{\mu}(1720)\,\, \right), 
\label{klan1720}\\
{\cal L}_{K^*(892) \Lambda N} &=& - g_{K^*(892) \Lambda N} 
( \bar{N} \gamma^\mu  \Lambda K^*_\mu(892) + 
\bar{K^*}_\mu(892) \bar{\Lambda}  \gamma^\mu N ),
\label{kslan}\\
{\cal L}_{K^*(892) K \pi} &=& i f_{K^*(892) K \pi} 
\left( \bar{K} \vec\tau K^*_\mu(892) \cdot 
\partial^\mu \vec\phi
- (\partial^\mu \bar{K}) \vec\tau K^*_\mu(892) \cdot \vec\phi \,\right) 
+ {\rm h. c.}\,.\qquad
\label{kskpi}
\end{eqnarray}
The amplitudes are given by: 
${\cal M}_{\pi^0 p \rightarrow \Lambda K^+}  
= -{\cal M}_{\pi^0 n \rightarrow \Lambda K^0}
= \frac{1}{\sqrt{2}}{\cal M}_{\pi^+ n \rightarrow \Lambda K^+} 
= \frac{1}{\sqrt{2}}{\cal M}_{\pi^- p \rightarrow \Lambda K^0}
= {\cal M}_{a} + {\cal M}_{b} + {\cal M}_{c} + {\cal M}_{d}$,  
where the amplitudes ${\cal M}_{a}$, ${\cal M}_{b}$, ${\cal M}_{c}$
and ${\cal M}_{d}$ correspond to the diagrams 
(a), (b), (c) and (d), respectively in Fig. 1.

To carry the calculations further, form factors
are introduced which reflect the finite size of the hadrons. 
Those form factors
are carried by each vertex.
For the meson-baryon-(baryon resonance) vertex, the following form 
factor is used:
\begin{equation}
F({\vec q})=\frac{\Lambda_C^2}{\Lambda_C^2 + \vec{q}\,^2},
\label{form1}
\end{equation}
where $\vec{q}$ is the meson momentum, 
and  $\Lambda_C$ is the cut-off parameter. 
On the other hand, for the $K^*(892)$-$K$-$\pi$ vertex, the  form factor 
studied in Ref. \cite{gob} is used:
\begin{equation}
F_{K^*(892) K \pi}(\mid \frac{1}{2}(\vec{p_K}-\vec{p_\pi}) \mid) 
= C \mid \frac{1}{2}(\vec{p_K}-\vec{p_\pi}) \mid 
\exp\left( - \beta \mid\frac{1}{2}(\vec{p_K}-\vec{p_\pi})\mid^2 \right). 
\label{form2}
\end{equation}
Before discussing the results, the model parameters 
need to be specified.
The cut-off parameter $\Lambda_C$ appearing in Eq. (\ref{form1}) is
$\Lambda_C=0.8$ GeV for all meson-baryon-(baron resonance) vertices.
The values obtained for the coupling constants 
with this cut-off value are given in Table 1.
The fitted value for $g_{K^*(892)\Lambda N}$ 
is $g_{K^*(892)\Lambda N}$ = 0.45.
The other parameters $C$ and $\beta$ 
appearing in Eq. (\ref{form2}) are  $C = 2.72$ fm and 
$\beta = 8.88 \times 10^{-3}$ fm$^2$ obtained in Ref. \cite{gob}.

Here, it is appropriate to discuss $K^*(892)$ exchange.
Our calculations were also performed with the inclusion of the tensor coupling
interaction. However, it was found that the results show 
similar dependence on both c.m. energy and angle (or $\cos \theta_{c.m.}$) 
to those results calculated with 
the inclusion of the vector coupling interaction alone.
Thus, for the present purposes it is enough to include only the vector 
coupling interaction.

The energy dependence of the total cross sections
$\pi^- p \rightarrow \Lambda K^0$ 
is shown in Fig. 2. Figures (a) and (b) correspond to cases without and 
with the inclusion of interference terms, respectively.
The sign combination among the interference terms 
in calculation (b) is selected in such a way 
that {\it both} total and differential cross sections are
reproduced simultaneously.
Mere inclusion of the single resonance 
or the $K^*$ exchange alone cannot reproduce
the energy dependence of the total cross section data.

Next, the differential cross sections  
$\pi^- p \rightarrow \Lambda K^0$
are shown in Fig. 3.
Figures (a), (b) and (c) correspond to the
pion beam momenta 0.980 GeV/c, 1.13 GeV/c and 1.455 GeV/c,
respectively.
The general trends are reproduced, but the details are not yet satisfactory.

In Fig. 4, we give the energy dependence of the total cross sections  
$\pi N \rightarrow \Sigma K$.
In Ref. \cite{tsu1}, the $\Delta(1920)$ resonance 
was treated as an effective resonance which simulates 
other $\Delta$ resonance effects around the mass region 1.9 GeV.
The two coupling constants 
$g_{K \Sigma \Delta(1920)}$ and $g_{\pi N \Delta(1920)}$
were scaled in Ref. \cite{tsu1}. 
The results obtained by using these scaled coupling
constants are denoted by {\it set 1}.
A more quantitative discussion of this scaling will be made below.

The experimental data \cite{data} show that 
there are six $\Delta$ resonances which make 
contributions to the
$\pi N \rightarrow \Sigma K$
reactions around the mass region 1.9 GeV. (See Table 2.)
In order to understand the scaling factor quantitatively, 
we compare:
$$
\frac{{\rm the~ contribution~ of~} \Delta(1920) {\rm ~to~} 
\pi N \rightarrow \Sigma K}
{{\rm all~ } \Delta{\rm s'(masses~around~1.9~GeV)~
contribution~ to~} \pi N \rightarrow \Sigma K} 
= \frac {10.4}{37.37}=0.278,
$$
$$
\sqrt{ \frac{g^2_{K\Sigma \Delta(1920)} g^2_{\pi N \Delta(1920)}    
( {\rm from~ branching~ ratio}) }{g^2_{K\Sigma \Delta(1920)}
g^2_{\pi N \Delta(1920)}({\rm scaled})} }
= \sqrt{\frac{1.11\times 0.417}{3.83\times 1.44}} = 0.289.
$$
This comparison shows that the scaling 
factor (= 1.861) for each coupling constant
$g_{K \Sigma \Delta(1920)}$ and $g_{\pi N \Delta(1920)}$ 
is consistent with the total branching ratio obtained by summing these 
$\Delta$ resonance contributions.  
However, because of this effective description, the differential cross 
sections cannot be reproduced well.
A more accurate determination of the branching ratios for these 
$\Delta$ resonances is necessary.

Here, the
$\pi \Delta \rightarrow Y K$
reactions should also be mentioned.
The parametrizations and figures
given in Ref. \cite{tsu2} were obtained by using the lower values of the 
branching ratios for the resonances to decay to $\pi \Delta$.
The parametrizations obtained by using the averaged
values as given in Tables 1 and 2 of Ref. \cite{tsu2} will
be given later. The difference between the previous parametrizations 
and those to be given later is that in the latter 
the change in the multiplication factor is large. 
However, the second term of the parametrization for the 
$\pi^+ \Delta^0 \rightarrow \Sigma^0 K^+$  
total cross section remains the same as before.
As for the differential cross sections
$\pi \Delta \rightarrow Y K$, 
they are almost constant
as a function of $cos\theta$ in the c.m. frame for the 
beam energies for which calculations were made.

Finally, the parametrizations of the total
cross sections sufficient for all channels  of the
$\pi B \rightarrow Y K$
reactions in units of mb are given by:
\begin{eqnarray*}
\sigma(\pi^- p \rightarrow \Lambda K^0) 
 &=& \frac{0.007665 (\sqrt{s}-1.613)^{0.1341}}{(\sqrt{s}-1.720)^2+0.007826},\\
\sigma(\pi^+ p \rightarrow \Sigma^+ K^+) 
 &=& \frac{0.03591 (\sqrt{s}-1.688)^{0.9541}}{(\sqrt{s}-1.890)^2+0.01548}
 + \frac{0.1594 (\sqrt{s}-1.688)^{0.01056}}{(\sqrt{s}-3.000)^2+0.9412},\\
\sigma(\pi^- p \rightarrow \Sigma^- K^+) 
 &=& \frac{0.009803 (\sqrt{s}-1.688)^{0.6021}}{(\sqrt{s}-1.742)^2+0.006583}
 + \frac{0.006521 (\sqrt{s}-1.688)^{1.4728}}{(\sqrt{s}-1.940)^2+0.006248},\\
\sigma(\pi^+ n \rightarrow \Sigma^0 K^+) 
&=& \sigma(\pi^0 n \rightarrow \Sigma^- K^+)
= \frac{0.05014 (\sqrt{s}-1.688)^{1.2878}}{(\sqrt{s}-1.730)^2+0.006455},\\
\sigma(\pi^0 p \rightarrow \Sigma^0 K^+) 
 &=& \frac{0.003978 (\sqrt{s}-1.688)^{0.5848}}{(\sqrt{s}-1.740)^2+0.006670}
 + \frac{0.04709 (\sqrt{s}-1.688)^{2.1650}}{(\sqrt{s}-1.905)^2+0.006358},\\
\sigma(\pi^- \Delta^{++} \rightarrow \Lambda K^+) 
 &=& \frac{0.009883 (\sqrt{s}-1.613)^{0.7866}}{(\sqrt{s}-1.720)^2+0.004852},\\
\sigma(\pi^- \Delta^{++} \rightarrow \Sigma^0 K^+) 
&=& \frac{0.007448 (\sqrt{s}-1.688)^{0.7785}}{(\sqrt{s}-1.725)^2+0.008147},\\
\sigma(\pi^0 \Delta^0 \rightarrow \Sigma^- K^+) 
&=& \frac{0.01052 (\sqrt{s}-1.688)^{0.8140}}{(\sqrt{s}-1.725)^2+0.007713},\\
\sigma(\pi^+ \Delta^0 \rightarrow \Sigma^0 K^+) 
&=& \frac{0.003100 (\sqrt{s}-1.688)^{0.9853}}{(\sqrt{s}-1.725)^2+0.005414}
+ \frac{0.3179 (\sqrt{s}-1.688)^{0.9025}}{(\sqrt{s}-2.675)^2+44.88},\\
\sigma(\pi^+ \Delta^- \rightarrow \Sigma^- K^+) 
&=& \frac{0.02629 (\sqrt{s}-1.688)^{1.2078}}{(\sqrt{s}-1.725)^2+0.003777},
\end{eqnarray*}
where all the parametrizations given above should be understood to be zero
below threshold.
Furthermore, the parametrizations for the
$\pi \Delta \rightarrow Y K$ 
reactions are obtained without the inclusion of interference terms.
The above are the completion of the parametrizations for the 
total cross sections
$\pi B \rightarrow Y K$.
Parametrizations for the other channels can be obtained by multiplying
the relevant constant factors arising from isospin space \cite{tsu1,tsu2}.

The next task is to investigate the
$B_1 B_2 \rightarrow B_3 Y K$
reactions by using the same resonance model, where most of the
model parameters have already been fixed by this investigation.
This program is now in progress.
\vspace{0.5cm}

\noindent {\bf Acknowledgement:} One of the authors (K.T.) expresses his 
thanks to Prof. A.W. Thomas and to  Drs. A.G. Williams and K. Saito
who organized $\lq\lq$Quarks, Hadron and Nuclei Workshop" held in
Adelaide for the opportunity to report the results mentioned 
here.


\vspace{0.5cm}
 
\noindent
{\bf Captions}
 
\noindent
{\bf Fig. 1: }
The processes contributing to the $\pi N \rightarrow \Lambda K$
reactions. The diagrams correspond to 
$(a):\, N(1650)\, I(J^P)=\frac{1}{2}(\frac{1}{2}^-)$, 
$(b):\, N(1710)\, \frac{1}{2}(\frac{1}{2}^+)$, 
$(c):\, N(1720)\, \frac{1}{2}(\frac{3}{2}^+)$ s-channels and
$(d):\, K^*(892)$-exchange, respectively.\\
\noindent
{\bf Fig. 2: }
The total cross sections
$\pi^- p \rightarrow \Lambda K^0$.
The experimental data are taken from Ref. \cite{bal}.
The results are: (a)-without and 
(b)-with the inclusion of interference terms, respectively.\\
\noindent
{\bf Fig. 3: }
The differential cross sections 
$\pi^- p \rightarrow \Lambda K^0$ 
in c.m. frame.
(a), (b) and (c) correspond to the pion beam momenta 
0.980 GeV/c ($\sqrt s = 1.66$ GeV), 1.13 GeV/c 
($\sqrt s = 1.742$ GeV) 
and 1.455 GeV/c ($\sqrt s =1.908$ GeV), respectively. 
For (a) and (b), the experimental data are taken from Ref. \cite{kna75},  
while for (c) they are taken from Ref. \cite{sax80}.\\
\noindent
{\bf Fig. 4: }
The total sections for the reactions 
(a): $\pi^+ p \rightarrow \Sigma^+ K^+$, 
(b): $\pi^- p \rightarrow \Sigma^- K^+$ and 
(c): $\pi^+ n \rightarrow \Sigma^0 K^+$ and 
$\pi^0 n \rightarrow \Sigma^- K^+$, respectively.
The notation {\it set 1} in these figures indicates 
the results obtained by using the
scaled values for 
$g_{K \Sigma \Delta(1920)}$ and $g_{\pi N \Delta(1920)}$.
For further explanations, see Ref. \cite{tsu1}.
\newpage
\begin{table}[tbp]
\caption{The calculated coupling constants and 
the experimental branching ratios.}
\begin{center}
\begin{tabular}{cccccccc}
\hline
$B^*$(resonance) & $\Gamma^{full}(MeV)$ & $\Gamma_{N \pi}(\%) $ 
& $g_{\pi N B^*}^2$ & $\Gamma_{\Lambda K}(\%)$ & $ g_{K \Lambda  B^*}^2 $ \\
\hline \\
$N(1650)$       & 150 & 70.0 & 1.41    & 7.0     & 6.40 $\times 10^{-1}$ \\
$N(1710)$       & 100 & 15.0 & 2.57    & 15.0    & 4.74 $\times 10^{+1}$ \\
$N(1720)$       & 150 & 15.0 & 5.27 $\times 10^{-2}$ & 6.5 & 3.91 \\
\hline
                & $f_{K^*(892) K \pi}^2$ & & $g_{K^*(892) \Lambda N }^2$ & \\
\hline
                & 6.89 $\times 10^{-1}$ &    & 2.03 $\times 10^{-1}$ & &   \\
                &($ \Gamma$=50 MeV,&$\Gamma_{K \pi}=100\% $) &   & & \\ 
\hline
\end{tabular}
\end{center}
\end{table}
%
\begin{table}[tbp]
\caption{Contribution of $\Delta (1920)$ to the $\pi N\rightarrow \Sigma K $
reactions compared with other possible $\Delta$ resonances between
1900 and 1940 MeV \protect\cite{data}.
The values in column four are all upper values.}
\begin{center}
\begin{tabular}{cccc}
\hline
$\Delta^*$(resonance) & $\Gamma_{total}(MeV)$ &  
${\bf (\Gamma_{\pi N \Delta^*}\Gamma_{K \Sigma \Delta^*})^{1/2}/\Gamma_{total}}$ & 
${\bf (\Gamma_{\pi N\Delta^*}\Gamma_{K \Sigma \Delta^*})^{1/2}} (MeV)$   \\
\hline \\
$\Delta(1900)$       & 200    & $< 0.03 $         &  6.00  \\
$\Delta(1905)$       & 350    & $\mid 0.015 \mid$ &  5.25  \\
$\Delta(1910)$       & 250    & $< 0.03  $        &  7.50  \\
$\Delta(1920)$       & 200    & $ \mid0.052 \mid $& 10.4   \\
$\Delta(1930)$       & 350    & $ <0.015$         & 5.25   \\
$\Delta(1940)$       & 198.4  & $ <0.015   $      & 2.98   \\
\hline  
\\
\end{tabular}
\end{center}
\end{table}
\end{document}